\documentclass{elsart}
\usepackage{xspace,amsmath,amssymb,amsfonts}

\usepackage[dvips]{graphicx}
\usepackage{bm}
\usepackage{amstext}

\newcommand{\dd}{{\rm d}}

\newtheorem{theorem}{Theorem}[section]
\newtheorem{corollary}[theorem]{Corollary}
\newtheorem{lemma}[theorem]{Lemma}

\theoremstyle{definition}

\theoremstyle{remark}
\newtheorem{remark}[theorem]{Remark}

\begin{document}
\runauthor{E. Minguzzi}
\begin{frontmatter}
\title{Characterization of some causality conditions through the continuity of the Lorentzian distance}
\author[E. Minguzzi]{Ettore Minguzzi\thanksref{a}}
\address[E. Minguzzi]{Dipartimento di Matematica Applicata, Universit\`a degli Studi di Firenze,  Via
S. Marta 3,  I-50139 Firenze}
\thanks[a]{{\it E-mail address:} {\tt ettore.minguzzi@unifi.it}}
\begin{abstract}
A classical result in Lorentzian geometry states that  a strongly
causal spacetime is globally hyperbolic if and only if the
Lorentzian distance is finite valued for every metric choice in the
conformal class. It is proved here that a non-total imprisoning
spacetime is globally hyperbolic if and only if for every  metric
choice in the conformal class the Lorentzian distance is continuous.
Moreover, it is proved that a non-total imprisoning spacetime is
causally simple  if and only if for every  metric choice in the
conformal class the Lorentzian distance is continuous wherever it
vanishes. Finally, a strongly causal spacetime is causally
continuous if and only if there is at least one metric in the
conformal class such that the Lorentzian distance is continuous
wherever it vanishes.

\end{abstract}
\end{frontmatter}

\section{Introduction}
Let $(M,g)$ be a spacetime, the Lorentzian distance function $d:
M\times M \to [0,+\infty]$ is defined by,
\[
d(p,q)=\textrm{sup}_{\gamma} \,l(\gamma)
\]
where $\gamma:[a,b] \to M$ is the generic $C^1$ causal curve
connecting $p$ to $q$ and $l$ is the Lorentzian length functional
$l(\gamma)=\int_a^b \sqrt{-g(\dot{\gamma},\dot{\gamma})} \, \dd t$.
If there is no causal curve connecting $p$ to $q$ then it is
understood that $d(p,q)=0$.

The Lorentzian distance is clearly a non-conformally invariant
concept as the Lorentzian length changes under multiplication of the
metric by a positive function. However, given the function $d$ the
chronological relation is determined through the equivalence $(x,z)
\in I^{+}\Leftrightarrow d(x,z)>0$.  It must therefore be possible
to relate more or less directly some properties of the Lorentzian
distance with the causality properties of spacetime.

As a matter of fact there can be more than one characterization of a
given causality property. For instance, future  distinction can be
characterized as \cite[Lemma 4.23]{beem96}
\begin{center}
 for each pair of distinct $p,q \in M$, there is some $x \in M$, \\
such that exactly one of $d(p,x)$ and $d(q,x)$ is zero,
\end{center}
which is merely a restatement of the original definition, $x\ne z
\Rightarrow I^{+}(x) \ne I^{+}(z)$, taking into account the
equivalence $(x,z) \in I^{+}\Leftrightarrow d(x,z)>0$.
Alternatively, there is the possibility of characterizing a
spacetime as future distinguishing through the following theorem
which is analogous to \cite[Theorem 4.27]{beem96} for strong
causality.

\begin{theorem} \label{njs}
 A spacetime is future distinguishing iff every point $x\in M$ admits arbitrary small neighborhoods $V$  such that the
restricted distance $d(x,\cdot)\vert_V: V \to [0,+\infty]$ coincides
with  $d_V(x, \cdot)$ where $d_V$ is the Lorentzian distance of
spacetime $(V,g\vert_V)$. (Moreover, in this case $V$ can be chosen
globally hyperbolic, so that $d(x,\cdot)\vert_V$ is actually finite
and continuous.) An analogous past version holds.
\end{theorem}

(the proof is postponed to the end of the introduction)


%
Characterizations of the last type should be preferred because,
although sometimes more difficult to obtain, they provide new
information on some non-trivial features of the Lorentzian distance,
namely its finiteness, continuity and locality properties.

The Lorentzian distance has been successfully used to characterize
in this way some causality properties among which the most important
are strong causality and global hyperbolicity \cite{beem96}. In the
literature no characterization can be found of causal continuity and
causal simplicity, a gap which will be filled by this work.


I refer the reader to \cite{minguzzi06c} for most of the conventions
used in this work. In particular, I denote with $(M,g)$ a $C^{r}$
spacetime (connected, time-oriented Lorentzian manifold), $r\in \{3,
\dots, \infty\}$ of arbitrary dimension $n\geq 2$ and signature
$(-,+,\dots,+)$. On $M\times M$ the usual product topology is
defined. For convenience and generality I often use the causal
relations on $M \times M$ in place of the more widespread point
based relations $I^{+}(x)$, $J^{+}(x)$, $E^{+}(x)$ (and past
versions). The subset symbol $\subset$ is reflexive, $X \subset X$.

With $(M, \bm{g})$ it is denoted the conformal structure namely the
class of spacetimes which share the same time orientation of a
representative $(M,g)$ but for which the metric may differ from $g$
by a positive conformal factor $g'=\Omega(x) g$, $\Omega
>0$. With the boldface notation $\bm{g}$ it is also denoted the set
of metrics conformal to $g$. I shall write ``spacetime
$(M,\bm{g})$'' by meaning with this  the conformal structure.

A classical result  by Beem and Ehrlich \cite[Theorem 3.5]{beem79},
\cite[Theorem 4.30]{beem96}, states that a strongly causal spacetime
$(M,\bm{g})$ is globally hyperbolic if and only if for every $g \in
\bm{g}$, $d_g: M\times M \to [0,+\infty]$ is finite valued where
$d_g$ is the Lorentzian distance of $(M,g)$. In this case the
Lorentzian distance is also continuous \cite[Lemma 4.5]{beem96}, and
moreover, whenever $(p,q) \in J^{+}$, it is maximized by a suitable
connecting geodesic $\eta$, $d(p,q)=l(\eta)$. These good properties
are lost even by considering the property of causal simplicity which
stays just below globally hyperbolicity in the causal ladder of
spacetimes \cite{hawking74,minguzzi07c}. Nevertheless, I shall prove
in this work that both causal simplicity and causal continuity admit
a characterization through the continuity properties of the
Lorentzian distance function. Moreover, the continuity of the
Lorentzian distance in all the conformal class can also be used, in
the same way as the finiteness property, to characterize global
hyperbolicity (see theorem \ref{nod}).

Recall that a spacetime is {\em causally simple} if
\cite{bernal06b,minguzzi07c} it is causal and $\bar{J}^{+}=J^{+}$,
while it is {\em causally continuous} if \cite{minguzzi07e} it is
{\em weakly distinguishing}  (i.e. $I^{+}(x)=I^{+}(y)$ and
$I^{-}(x)=I^{-}(y) \Rightarrow x=y$) and {\em reflective} (i.e.
$I^{+}(y) \subset I^{+}(x) \Leftrightarrow I^{-}(x) \subset
I^{-}(y)$). A spacetime is {\em future distinguishing} if
$I^{+}(x)=I^{+}(y) \Rightarrow x=y$. Analogously, it is {\em past
distinguishing} if $I^{-}(x)=I^{-}(y) \Rightarrow x=y$. A spacetime
is {\em distinguishing} if it is both future and past
distinguishing. A spacetime is {\em non-total imprisoning} if no
future inextendible causal curve is contained in a compact
(replacing {\em future} with {\em past} gives the same property
\cite{beem76,minguzzi07f}). A spacetime is {\em non-partial
imprisoning} if there is no inextendible causal curve which returns,
in the past or future direction, indefinitely into a compact.

It will be useful to keep in mind the chain of implications: global
hyperbolicity $\Rightarrow$ causal simplicity $\Rightarrow$ causal
continuity $\Rightarrow$ stable causality $\Rightarrow$ strong
causality $\Rightarrow$ non-partial imprisonment $\Rightarrow$
distinction $\Rightarrow$ past or future distinction $\Rightarrow$
weak distinction $\Rightarrow$ non-total imprisonment $\Rightarrow$
causality (see \cite{hawking73,minguzzi06c,minguzzi07f}). Finally,
recall that a set is an {\em arbitrarily small} neighborhood of $x
\in M$, if it can be chosen to be contained in any given
neighborhood of $x$. The reader will be assumed to be familiar with
 the limit curve theorem  given in
\cite{minguzzi07c} (which generalizes and strengthens the limit
curve theorem given in \cite{beem96}). This is the only limit curve
theorem to which we shall make reference throughout the paper.

We end the section by giving the proof to theorem \ref{njs}.

\begin{pf}[of theorem \ref{njs}]\\
$\Rightarrow$. In this direction the proof has been given by the
author in \cite[Theorem 2.12]{minguzzi06f}. It is included here for
completeness.

Since $M$ is future distinguishing \cite[Lemma 3.10]{minguzzi06c}
for every open set $U\ni x$ there is a neighborhood $V \subset U$,
$V\ni x$ such that every causal curve starting from $x$ and ending
at $y \in V$, is necessarily contained in $V$ (this is stated also
in \cite[Sect. 6.4]{hawking73}).  As a consequence $d(x,
\cdot)\vert_V: V \to [0,+\infty]$ coincides with $d_{V}(x,\cdot)$,
where $ d_{V}$ is the Lorentzian distance on the spacetime
$(V,g\vert_{V})$.

Moreover, the same proof \cite[Lemma 3.10]{minguzzi06c} shows that
in this case $V$ can be chosen globally hyperbolic when regarded as
a spacetime with the induced metric. Since $V$ is globally
hyperbolic $d_{V}(x,\cdot)$ is continuous and finite and so is $d(x,
\cdot)\vert_V: V \to [0,+\infty]$.

$\Leftarrow$. It is obvious that for every point $x$ and open
neighborhood $V\ni x$, $I^{+}(x,V)\subset I^{+}(x)\cap V$. The
converse is true provided we choose $V$ so that $d_{V}(x,\cdot)=d(x,
\cdot)\vert_V$ because if $y \in I^{+}(x)\cap V$ then $d(x, y)>0$
which implies $d_V(x,y)>0$ and hence $y \in I^{+}(x,V)$. Thus for
every $x$ there is an arbitrary small open set $V$ such that
$I^{+}(x,V)=I^{+}(x)\cap V$. This property characterizes future
distinction, see \cite[Lemma 3.10 and Remark 3.12]{minguzzi06c}.
\hfill$\square$
\end{pf}

\section{Continuity on the vanishing distance set}

The Lorentzian distance vanishes on the {\em vanishing distance set}
$(M\times M)\backslash I^{+}$. This set is clearly conformally
invariant, despite the fact that the Lorentzian distance is not. The
idea is to use the continuity properties of  the Lorentzian distance
on $(M\times M)\backslash I^{+}$ to characterize causal simplicity
and causal continuity. Note that the Lorentzian distance vanishes on
the {\em open} set $(M\times M)\backslash \bar{I}^{+}$ thus it is
there continuous. Hence the Lorentzian distance is continuous on the
vanishing distance set if and only if it is continuous on
$\dot{I}^{+}=\dot{J}^{+}$ (for a proof of this equality and
$\bar{I}^{+}=\bar{J}^{+}$ see \cite{minguzzi06c}).


\begin{lemma} \label{lom}
Let $(M,\bm{g})$ be a non-total imprisoning spacetime and let $(x,z)
\in \bar{J}^{+}\backslash J^{+}$.  There is a representative of the
conformal class such that the Lorentzian distance  is not finite in
any neighborhood of $(x,z)$, in particular it has an infinite
discontinuity at $(x,z)$ where the Lorentzian distance vanishes.
\end{lemma}

\begin{pf} Let $(M,g)$ be any representative in the conformal class. Let $\gamma_n$ be a sequence of timelike
curves with endpoints $(x_n,z_n) \to (x,z)$. There is no compact set
which contains an infinite subsequence of $\gamma_n$ (otherwise by
non-total imprisonment and according to the limit curve theorem
\cite{minguzzi07c} there would be a limit continuous causal curve
connecting $x$ to $z$ which is impossible because $(x,z) \notin
J^{+}$) thus for every compact set all but a finite number of
$\gamma_n$ are not contained in the compact set. Let $h$ be an
auxiliary complete Riemannian metric on $M$, let $\rho$ be the
associated distance, and let $B_n(x)$ be the open balls centered at
$x$ and of radius $n$ with respect to $h$. It is possible to pass to
a subsequence, denoted again $\gamma_n$, such that there is $p_n \in
\gamma_n \cap [M\backslash B_n(x)]$.
 Let $\Omega_n\ge 1$ be
a smooth function equal to 1 outside $A_n = \{r \in M:
n-1<\rho(x,r)<n\}$ and sufficiently large in $A_n$ so that the
Lorentzian length with respect to the metric $\Omega_n g$ of the
timelike segment of $\gamma_n$ connecting $x_n$ to $p_n$  is larger
than $n$. Let $y \in I^{-}(x)$, $w \in I^{+}(z)$, so that  for
sufficiently large $n$, $y\ll x_n\ll p_n$, $d_{\Omega_n g}(y,p_n)\ge
d_{\Omega_n g}(x_n,p_n)>n$. Defined $\Omega=\Pi_{n} \Omega_n$, it
holds $d_{\Omega g} (y,p_n)>n$ for sufficiently large $n$. Since for
sufficiently large $n$, $y \ll p_n\ll w$,
\[
d_{\Omega g}(y,w)\ge d_{\Omega g}(y,p_n)+d_{\Omega g}(p_n,w)\ge
d_{\Omega g}(y,p_n)>n
\]
thus $d_{\Omega g}(y,w)=+\infty$ and since $y$ and $w$ can be chosen
arbitrarily close to $x$ and $z$, where $d_{\Omega g}(x,z)=0$ as
$(x,z) \notin J^{+}$, there follows the infinite discontinuity of
$d_{\Omega g}$ at $(x,z)$. \hfill$\square$
\end{pf}

\begin{theorem} \label{po1} The non-total imprisoning spacetime $(M,\bm{g})$ is causally
simple if and only if for every metric $g$ in the conformal class
the Lorentzian distance  is continuous on the vanishing distance
set.
\end{theorem}

\begin{pf}
$\Rightarrow$. In this direction the proof has been given by the
author in \cite[Theorem 3.10]{minguzzi06f}. It is  included here for
completeness.

Assume that $(M,g)$ is causally simple  and let $(x,z)\in
\dot{I}^{+}$ so that $d(x,z)=0$. If $(x,z)$ is a discontinuity point
for $d$, then there is a $\epsilon>0$ and a sequence $(x_n,z_n) \to
(x,z)$, such that $d(x_n,z_n)>\epsilon>0$. In particular $(x_n,z_n)
\in I^{+}$ and $(x,z) \in \bar{I}^{+}\backslash
I^{+}=\dot{I}^{+}=E^{+}$, by causal simplicity \cite[Lemma
3.67]{minguzzi06c}. Let $\gamma_n$ be causal curves connecting $x_n$
to $z_n$ and such that $\limsup_{n \to +\infty} l(\gamma_n)
\ge\epsilon$ (for instance let $l(\gamma_n)> d(x_n,z_n)-\frac{1}{n}$
if $d(x_n,z_n)<+\infty$ and $l(\gamma_n) >n$ if
$d(x_n,z_n)=+\infty$). By the limit curve theorem there is a
continuous causal curve $\gamma$ passing through $x$ and a
distinguishing subsequence $\gamma_j$ which converges to it. But by
construction any event $y\ne x,z,$ of $\gamma$ is the limit of
events $y_j \in \gamma_j$, $ (x_j, y_j) \in J^{+}$, hence $(x,y) \in
\bar{J}^{+}=J^{+}$ and analogously $(y,z) \in J^{+}$, thus $\gamma$
must be a lightlike geodesic connecting $x$ to $z$, otherwise $(x,z)
\in I^{+}$. Finally, by using the upper semi-continuity of the
length functional
\[
d(x,z) \ge l(\gamma) \ge \limsup_{j \to +\infty} l(\gamma_j)\ge
\epsilon>0.
\]
The contradiction proves that $(x,z)$ cannot be a discontinuity
point for the Lorentzian distance.

$\Leftarrow$. Since $(M,g)$ is non-total imprisoning it is causal.
Assume that $(M,g)$ is not causally simple, then since a causally
simple spacetime is a causal spacetime for which $\bar{J}^+=J^{+}$,
it must be $\bar{J}^+\ne J^{+}$, that is there is a pair $(x,z) \in
\bar{J}^{+}\backslash J^{+}$. The thesis is now a consequence of
lemma \ref{lom}. \hfill$\square$
\end{pf}

Recall that the {\em timelike diameter} of a spacetime $(M,g)$ is
defined by
\[
\textrm{diam}(M,g)=\sup\{d(p,q): p,q \in M\}
\]
that is, it is the least upper bound of the Lorentzian lengths of
the $C^1$ causal curves on spacetime.

\begin{lemma} \label{pha}
Let $h$ be an auxiliary complete Riemannian metric on $M$ and let
$\rho$ be the associated distance. Let $q \in M$ and let
$B_n(q)=\{r: \rho(q,r)<n\}$ be the open balls of radius $n$ centered
at $q$. If $(M,\bm{g})$ is strongly causal, then  there is a
representative $g$, such that $\textrm{diam}(M,g)$ is finite and for
every $\epsilon>0$ there is a $n \in \mathbb{N}$ such that if
$\gamma: I \to M$ is any $C^1$ causal curve,
\[
\int_{I\cap \gamma^{-1}(M\backslash \bar{B}_n) } \!\!\!\!\!\!\!
\sqrt{-g(\dot{\gamma},\dot{\gamma})} \, \dd t <\epsilon
\]
that is, its many connected pieces contained in the open set $
M\backslash \bar{B}_n$ have a total Lorentzian length less than
$\epsilon$.
\end{lemma}

\begin{pf}
Let $\tilde{g}$ be a representative in the conformal class. Every
point $p$ admits a causally convex neighborhood $U(p)$ such that its
closure is compact and contained in a globally hyperbolic
neighborhood $V \supset \bar{U}$. Since $U$ is causally convex in
$M$ the Lorentzian distance of spacetime $(U,\tilde{g}\vert_U)$
coincides with $\tilde{d}\vert_{U\times U}$ where $\tilde{d}$ is the
Lorentzian distance of $(M,\tilde{g})$. However, since $V \subset
M$, $U$ is also  causally convex in $V$, thus this same distance
coincides with the restriction of $\tilde{d}_{(V,\tilde{g}\vert_V)}$
to the set $U\times U$, but since $\bar{U}\times \bar{U} \subset V
\times V$ is compact and $\tilde{d}_{(V,\tilde{g}\vert_V)}$ is
continuous, $\tilde{d}\vert_{U\times U}$ is actually bounded and
continuous. Thus there is a constant that bounds the Lorentzian
length of all the causal curves contained in $U$.

Every compact set $W_n=\bar{B}_{n+2}(q) \backslash B_{n+1}(q)$ can
be covered with a finite number of such causally convex
neighborhoods $U(p_i)$ with compact closure contained in
$B_{n+3}(q) \backslash \bar{B}_{n}(q)$. Let $A_n=\cup_i U(p_i)$,
$\bar{A}_n \subset B_{n+3}(q) \backslash \bar{B}_{n}(q)$. Since
every causal curve $\eta$ can pass through $U(p_i)$ only once, and
the segment there contained is bounded by a constant depending on
the causally convex neighborhood, the length of (the many pieces of)
$\eta\cap W_n$ is bounded from above by a constant $k_n$ independent
of $\eta$. Let $\varepsilon>0$ be given and let $0<\Omega_n\le 1$ be
a conformal factor such that $\Omega_n=1$ outside $B_{n+3}(q)
\backslash \bar{B}_{n}(q)$ and sufficiently small on $\bar{A}_n$
that $k_n<\varepsilon/{2^n}$.

Defining $g=(\Pi_n \Omega_n) \tilde{g}$ the Lorentzian distance $d$
of $(M,g)$ is such that if $\gamma$ is a causal curve the length of
the many pieces of $\gamma \cap M\backslash B_{n+1}(q)$ is bounded
by $\sum^{+\infty}_{i=n} k_i=\varepsilon/2^{n-1}$. From this fact
the thesis follows. \hfill$\square$
\end{pf}

\begin{theorem} \label{po2}
The strongly causal spacetime $(M,\bm{g})$ is causally continuous if
and only if there is a metric $g$ in the conformal class  such that
the Lorentzian distance  is continuous on the vanishing distance
set.
\end{theorem}

\begin{pf}
$\Leftarrow$.  In this direction  the proof has been given by the
author in \cite[Corollary 3.4]{minguzzi06f}. Actually it suffices to
assume weak (or even feeble) distinction instead of strong
causality. I repeat the proof here for completeness. Assume $(M,g)$
has a continuous Lorentzian distance on the vanishing distance set.
Note that we have only to prove that $(M,g)$ is reflective. If
$(M,g)$ were not reflective then it would be non past or non future
reflective. We can assume the first possibility as the other case
can be treated similarly. Thus there is a pair $(x,z)$ and an event
$y$ such that $I^{+}(x) \supset I^{+}(z)$ but $y \in I^{-}(x)$ while
$y \notin I^{-}(z)$. In particular, $d(y,z)=0$. Since
$I^{+}(z)\subset I^{+}(x)$, $z \in \bar{I}^{+}(x)$. Let $z_n \to z$,
$z_{n} \in I^{+}(x)$, then
\[
d(y,z_n)\ge d(y,x)+d(x,z_{n})>d(y,x)>0,
\]
thus there is a discontinuity at $(y,z)$, where $d(y,z)=0$, a
contradiction.

$\Rightarrow$. Let $(M,\bm{g})$ be causally continuous and let $q
\in M$. Consider the representative $(M,g)$ in the conformal class
with the properties mentioned in the statement of lemma \ref{pha}
and let $h$ be the complete Riemannian metric there mentioned. We
have to show that $d$ is continuous at $(x,z) \in \dot{I}^{+}$, that
is if  $(x_k,z_k) \to (x,z)$ then for any given $\epsilon>0$, for
sufficiently large $k$, $d(x_k,z_k)\le \epsilon$. If it were not
then there would be a $\epsilon>0$ and a subsequence $(x_n,z_n) \to
(x,z)$ such that $d(x_n,z_n)> \epsilon$. Since the timelike diameter
is finite $d(x_n,z_n)<+\infty$ and we can consider a limit
maximizing sequence \cite{beem96,minguzzi07c} of timelike curves
$\gamma_n$ connecting $x_n$ to $z_n$ with length
$l(\gamma_n)>\epsilon$ (limit maximizing means
$d(x_n,z_n)-l(\gamma_n)\to 0$). Note that $x \ne z$, indeed if $x=z$
since $l(\gamma_n)>\epsilon$ there is a neighborhood $U$ of $x$ such
that none of the curves $\gamma_n$ are contained in $U$ (the bound
to the Lorentzian length is a consequence, see \cite[proof of (b)
theorem 2.4]{minguzzi07c}, of the bound on the Riemannian length,
which goes to zero with the size of the neighborhood see \cite[Sect.
3.3]{beem96} \cite[Lemma 2.5]{minguzzi07c}). However, by strong
causality at $x$ this is impossible, thus it must be $x\ne z$.

Parametrize $\gamma_n$ with respect to $h$-length, so that they have
domain $[a_n,b_n]$, $x_n=\gamma_n(a_n)$, $z_n=\gamma_n(b_n)$.  We
are going to apply the limit curve theorem \cite[Theorem
3.1]{minguzzi07c} case 2. The limit causal curve cannot connect $x$
to $z$ for otherwise this causal curve would have length $\lim_{n\to
+\infty}l(\gamma_n)=\lim_{n\to +\infty}d(x_n,z_n)>\epsilon$  (see
\cite{beem96,minguzzi07c}) and hence $(x,z) \in I^{+}$ a
contradiction (thus subcase $b<+\infty$ of \cite[Theorem
3.1]{minguzzi07c} case 2 does not apply).

We are now going to prove that there is a subsequence (denoted in
the same way) such that for sufficiently large $n$, $l(\gamma_n)\le
\epsilon$ which again is a contradiction.

\begin{figure}[ht]
\centering
\includegraphics[width=8cm]{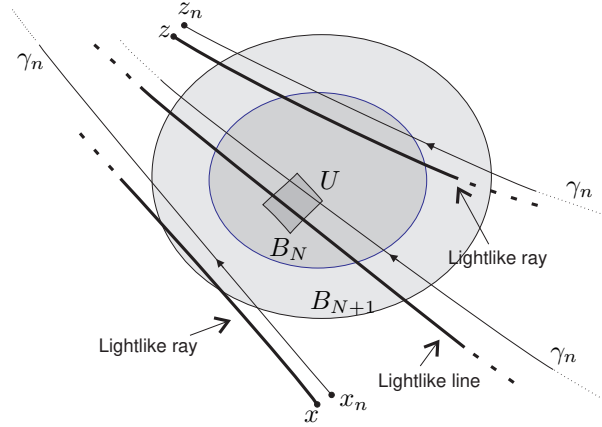} \caption{The sequence of causal curves $\gamma_n$ has only a  finite number of limit curves intersecting  the compact set $\bar{B}_N$.
Indeed, since the convergence is uniform on compact subsets to each
limit curve, given a causally convex neighborhood intersected by the
limit curve $U\subset B_{N+1}$, the limit sequence has to enter and
escape it. This can happen only a finite number of times (at most
once for each causally convex set U  of the covering of $\bar{B}_N$)
thus the limit curves are finite in number. Since the limit causal
curves are all lightlike geodesics, by the upper semi-continuity of
the length functional the contribution to the length of $\gamma_n$
coming from $B_{N+1}$ can be controlled.}
 \label{dim}
\end{figure}

Choose $N$ so that the compact set $\bar{B}_N(q)$ is such that the
Lorentzian length of $\gamma \cap M\backslash \bar{B}_N(q)$ is
smaller than $\epsilon/2$, where $\gamma$ is a generic causal curve.
Passing to a subsequence it is possible to assume that $\lim
(b_n-a_n)$ exists, in particular since subcase $b<+\infty$ of
\cite[Theorem 3.1]{minguzzi07c} case 2 does not apply it must be
$\lim (b_n-a_n)=+\infty$.  If all but a finite number of $\gamma_n$
do not intersect $\bar{B}_N(q)$, then there is nothing to prove.

More generally there will be some limit point in  $\bar{B}_N$ and
hence some limit curve (see figure \ref{dim}). Observe that every
limit curve must be either a lightlike line or a lightlike ray
(starting from $x$ or ending at $z$). Indeed, assume on the contrary
that the limit curve has two points $p \ll q$, then since $(x,p) \in
\bar{J}^{+}$ by future reflectivity $p \in \bar{J}^{+}(x)$,
analogously, $(q,z) \in \bar{J}^{+}$ and by past reflectivity $q \in
\bar{J}^{-}(z)$, finally, since $I^{+}$ is open $x \ll z$ a
contradiction.

Note that every limit curve, since the spacetime is non-partial
imprisoning, has to escape $\bar{B}_{N+1}$ to never reenter it.
Moreover,  let $\{U_1, \ldots, U_k\}$, $U_i \subset B_{N+1}$, be a
covering of $\bar{B}_N$ with causally convex subsets with compact
closures. The limit curve intersects and escapes at least one of
these sets and thus, for large $n$, the same is true for the curves
$\gamma_n$, which converge to the limit curve uniformly on compact
subsets (with respect to a complete Riemannian metric). Thus it is
possible to find a subsequence (denoted in the same way) which has
at most $k$ limit causal curves passing through $\bar{B}_N$ and such
that all the limit points  belong to one of these curves. Since the
convergence is uniform on compact subsets, the length functional is
upper semi-continuous and the limit curves are all lightlike, for
sufficiently large $n$ the contribution to the length of $\gamma_n$
coming from the open set $B_{N+1}$ is less than $\epsilon/2$ for
sufficiently large $n$. Moreover the length coming from the open set
$M\backslash \bar{B}_{N}$ is again $\epsilon/2$, thus
$l(\gamma_n)\le \epsilon$, for sufficiently large $n$, and the
thesis is proved. \hfill$\square$

\end{pf}

\section{Characterizations of global hyperbolicity}

In this section a new characterization of global hyperbolicity in
terms of the Lorentzian distance is obtained.

\begin{lemma}
In any spacetime the two properties
\begin{description}
\item{(i)} $\forall x,z$, $\overline{J^{+}(x) \cap J^{-}(z)}$ is compact,
\item{(ii)} $\forall x,z$, $\overline{I^{+}(x) \cap I^{-}(z)}$ is compact,
\end{description}
are equivalent, and implied by
\begin{description}
\item{(iii)} $\forall x,z$, $J^{+}(x) \cap J^{-}(z)$ is compact.
\end{description}
In a non-total imprisoning spacetime they are all  equivalent.
\end{lemma}

\begin{pf}
(i) $\Rightarrow$ (ii). Since $I^{+}(x) \subset J^{+}(x)$,
$I^{-}(z)\subset J^{-}(z)$, the closed set $\overline{I^{+}(x) \cap
I^{-}(z)}$ being a closed subset of the compact set
$\overline{J^{+}(x) \cap J^{-}(z)}$ is compact.\\
(ii) $\Rightarrow$ (i). Take $x' \ll x$, $z' \gg z$, then
$\overline{I^{+}(x') \cap I^{-}(z')}$ is compact. Since $J^{+}(x)
\subset I^{+}(x')$, $J^{-}(z')\subset I^{-}(z)$, the closed set
$\overline{J^{+}(x) \cap J^{-}(z)}$ being a closed subset of the
compact set $\overline{I^{+}(x') \cap I^{-}(z')}$ is compact. \\
(iii) $\Rightarrow$ (i). Trivial.\\
(i) $\Rightarrow$ (iii) in the non-total imprisoning case. Assume
$(M,g)$ is non-total imprisoning. We are going to prove that (i)
implies the closure of $J^{+}(y)$ and $J^{-}(y)$ for all $y$. From
that it follows for every $x,z$, $\overline{J^{+}(x) \cap
J^{-}(z)}=J^{+}(x) \cap J^{-}(z)$ and thus the compactness of this
last set. Assume for instance that $J^{+}(y)$ is not closed and let
$w \in \bar{J}^{+}(y) \backslash J^{+}(y)$. Take $r \gg w$, and a
sequence $r_n$, such that $w\ll r_n\ll r$, $r_n \to w$. Let
$\sigma_n$ be causal curves connecting $y$ with $r_n$. They are all
contained in the compact set $C=\overline{J^{+}(y)\cap J^{-}(r)}$.
By the limit curve theorem \cite{minguzzi07c} there is a limit
causal curve $\sigma$ starting from $y$, which necessarily joins $y$
to $w$, otherwise $\sigma$ would be future inextendible but
contained in the compact $C$, in contradiction with non-total
imprisonment. Thus $w \in J^{+}(y)$ again a contradiction which
proves that $\bar{J}^{+}(y)=J^{+}(y)$. \hfill$\square$
\end{pf}

\begin{remark}
In the previous lemma non-total imprisonment cannot be weakened to
causality, see Carter's example \cite[Fig. 39]{hawking73}.
\end{remark}

As a preliminary step we obtain this new characterization of global
hyperbolicity.

\begin{corollary} \label{bzx}
A spacetime is globally hyperbolic iff it is non-total imprisoning
and such that for every pair $x,z \in M$, $\overline{I^{+}(x) \cap
I^{-}(z)}$ is compact.
\end{corollary}

\begin{pf}
It suffices to recall that a spacetime is globally hyperbolic iff it
is causal and for every $x,z$, $J^{+}(x) \cap J^{-}(z)$ is compact
\cite{minguzzi06c,bernal06b}. \hfill$\square$
\end{pf}

\begin{remark}
Non-total imprisonment cannot be weakened to causality, see again
Carter's example \cite[Fig. 39]{hawking73}.
\end{remark}


\begin{lemma} \label{nos}
A causally simple spacetime is globally hyperbolic or it is possible
to find  events $x,z \in M$, $x \ll z$,  such that $J^{+}(x) \cap
J^{-}(z)$ is not compact and there is a sequence of causal curves
$\sigma_n: [0,a_n]\to M$ of endpoints $x$ and $z$, such that no
subsequence of $\sigma_n$ is contained in a compact set, $\sigma_n$
converges uniformly (with respect to a complete Riemannian metric)
on compact subsets to a future lightlike ray $\sigma^x$ starting
from $x$, and the reparametrized sequence
$\tilde{\sigma}_n(t)=\sigma_n(t-a_n)$, $\tilde{\sigma}_n:
[-a_n,0]\to M$ converges uniformly on compact subsets to a past
lightlike ray $\sigma^z$ ending at $z$.
\end{lemma}

\begin{pf}
Assume $(M,g)$ is not globally hyperbolic then since it is non-total
imprisoning there are $p\ll q$ such that $\overline{I^{+}(p) \cap
I^{-}(q)}$ is not compact thus since, $\overline{I^{+}(p) \cap
I^{-}(q)}\subset \bar{I}^{+}(p) \cap \bar{I}^{-}(q)=J^{+}(p) \cap
J^{-}(q)$, this last set is not compact. Let $\gamma: \mathbb{R} \to
M$ be a timelike curve connecting $p$ to $q$, such that
$p=\gamma(0)$, $q=\gamma(1)$. Let $B$ be the set of closed intervals
$[a,b]$, $a,b \in [0,1]$, such that $J^{+}(\gamma(a)) \cap
J^{-}(\gamma(b))$ is not compact (by causality $a <b$). The set $B$
is not empty because $[0,1] \in B$. Moreover, the set $B$ is ordered
by inclusion and we want to prove that it admits a minimal element.
By Hausdorff's maximum principle there is a maximal (totally
ordered) chain $\mathcal{C}$. The arbitrary intersection of convex
intervals is convex, thus the intersection of the elements in the
maximal chain, being the intersection of connected non-empty compact
intervals is a non-empty (this is a standard result in topology
\cite{engelking89}) connected compact interval
$[\tilde{a},\tilde{b}]$, $\tilde{a},\tilde{b} \in [0,1]$,
$\tilde{a}\le \tilde{b}$.

Actually, $\tilde{a} < \tilde{b}$ because of the following argument.
Assume $\tilde{a} = \tilde{b}$, the point $\gamma(\tilde{a})$ admits
a causally convex neighborhood $V$ with compact closure, and there
is $\epsilon>0$, such that the image of
$\gamma\vert_{[\tilde{a}-\epsilon, \tilde{a}+\epsilon]}$ is
contained in $V$. The interval $[\tilde{a},\tilde{a}+\epsilon]$
cannot belong to all the elements of the maximal chain, nor can the
interval $[\tilde{a}-\epsilon,\tilde{a}]$ thus there is an interval
$[a',b']$
 belonging to the maximal chain such that $[a',b'] \subset
 (\tilde{a}-\epsilon,\tilde{a}+\epsilon)$, thus $J^{+}(\gamma(\tilde{a}-\epsilon)) \cap
 J^{-}(\gamma(\tilde{a}+\epsilon))$ is not compact which is impossible because it is contained in $\bar{V}$. The
contradiction proves that $\tilde{a} < \tilde{b}$.

Define $x=\gamma(\tilde{a})$ and $z=\gamma(\tilde{b})$. In order to
prove the minimality of $[\tilde{a},\tilde{b}]$ on $B$ we have only
to show that $D=J^{+}(x) \cap
 J^{-}(z)$ is not compact, the minimality of $[\tilde{a},\tilde{b}]$ would follow trivially from the
 maximality of the chain $\mathcal{C}$.

Since $[\tilde{a},\tilde{b}]$ is the intersection of the elements of
$\mathcal{C}$, it is possible to construct (by arguing as above)
sequences $a_n \to a$, $b_n \to b$,  $a>a_{n+1}\ge a_n$,
$b<b_{n+1}\le b_n$, such that $J^{+}(\gamma(a_n))\cap
J^{-}(\gamma(b_n))$ is not compact.

Note that  $\bigcap_n J^{-}(\gamma(b_n)) \subset J^{-}(z)$ indeed,
if $w \in J^{-}(\gamma(b_n))$ for all $n$, then $\gamma(b_n) \in
J^{+}(w)$ and since $\gamma(b_n) \to z$, $z \in
\bar{J}^{+}(w)=J^{+}(w)$ which  implies $w \in J^{-}(z)$.
Analogously, $\bigcap_n J^{+}(\gamma(a_n)) \subset J^{+}(x)$. We
conclude
\begin{equation} \label{bid}
\bigcap_n\, [J^{+}(\gamma(a_n))\cap J^{-}(\gamma(b_n))] \subset
J^{+}(x)\cap J^{-}(z).
\end{equation}
Assume  that $D=J^{+}(x)\cap J^{-}(z)$ is  compact, let $h$ be a
complete Riemannian metric and let $B_n(p)$ be the open ball of
$h$-radius $n$ centered at $p$. Let $N$ such that $D \subset B_N(p)$
and let $E= \bar{B}_{N+1}(p)\backslash B_N(p)$ be a compact shell
which contains $D$ in its interior. Note that $E\cap
J^{+}(\gamma(a_n))\cap J^{-}(\gamma(b_n))$ is non-empty because the
causal curves issued from $\gamma(a_n)$ and reaching $\gamma(b_n)$
are not all included in a compact set (recall that the sets
$J^{+}(\gamma(a_{n}))\cap J^{-}(\gamma(b_{n}))$ are non compact) and
thus some of them cross $E$. The sets $E\cap J^{+}(\gamma(a_n))\cap
J^{-}(\gamma(b_n))$ give a nested family of non-empty compact
subsets whose intersection is, by Eq. (\ref{bid}), the empty set
which is impossible. The contradiction proves that $D$ is not
compact, and thus $[\tilde{a},\tilde{b}]$ is a minimal element for
$B$.

Let $\sigma_k$ be a sequence of causal curves not all contained in a
compact connecting $x$ to $z$. By the limit curve theorem
\cite{minguzzi07c} there is a subsequence $\sigma_n: [0,a_n] \to M$
and limit curves $\sigma^x$ and $\sigma^z$ as in the statement of
this theorem, but for their ` lightlike ray' nature which we have
still to prove. This is indeed a consequence of the minimality of
$[\tilde{a},\tilde{b}]$. Assume for instance that $\sigma^x$ is not
a lightlike ray, then there is $w \in \sigma^x \backslash\{x\}$,
such that $x \ll w$, and since $I^{+}$ is open there is $\delta>0$,
and $x'=\gamma(\tilde{a}+\delta)$, such that $(x',w) \in I^{+}$.
Since $w$ is a limit point for $\sigma_n$, it is possible to
construct a sequence of causal curves $\sigma_n'$ not entirely
contained in a compact, which connects $x'$ to $z$. This fact
implies that, $J^{+}(x')\cap J^{-}(z)$ is not  compact and hence
that $[\tilde{a}+\delta,\tilde{b}] \in B$ in contradiction with the
minimality of $[\tilde{a},\tilde{b}]$ in $B$. \hfill$\square$
\end{pf}

\begin{figure}[ht]
\centering
\includegraphics[width=6cm]{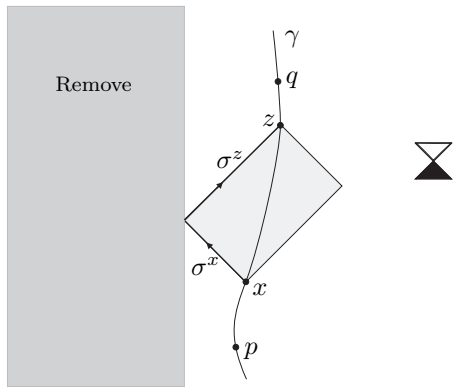} \caption{The argument of lemma \ref{nos}, which allows
us to construct lightlike rays $\sigma^x$ and $\sigma^z$ in a
causally simple non-globally hyperbolic spacetime (here given by 1+1
Minkowski spacetime with the usual coordinates $(t,x)$ restricted to
$x>0$).}
 \label{cons}
\end{figure}

\begin{theorem} \label{nod}
A non-total imprisoning spacetime $(M, \bm{g})$ is globally
hyperbolic if and only if for every metric choice in the conformal
class the Lorentzian distance is continuous.
\end{theorem}

\begin{pf}
$\Rightarrow$. This implications is well known, see \cite[Corollary
4.7]{beem96}. \\
$\Leftarrow$. For the converse, assume the spacetime $(M, \bm{g})$
is non-total imprisoning and that for every metric choice in the
conformal class the Lorentzian distance is continuous. By theorem
\ref{po1} the spacetime $(M, \bm{g})$ is causally simple. Let $h$ be
a complete Riemannian metric on $M$ and let $q \in M$. Take as
representative $g$ the metric selected by the statement of lemma
\ref{pha}, and let $B_n(q)$ be the open balls of $h$-radius $n$
centered at $q$. By lemma \ref{pha} $\textrm{diam}(M,g)$ is finite.
Assume that $(M,g)$ is not globally hyperbolic, then there are $x,z
\in M$, $x\ll z$, as in the statement of lemma \ref{nos}. Since
$\textrm{diam}(M,g)$ is finite, $d_g(x,z)<+\infty$. The idea is to
conformally change the metric outside the closed set $J^{+}(x)\cap
J^{-}(z)$. Any such change necessarily leaves unaltered the distance
between $x$ and $z$ but the conformal factor can be chosen so that
there is a infinite discontinuity for the new Lorentzian distance.
Let $z_n\gg z$ be a sequence of points such that $z_n \to z$, and
let $w_n\in \sigma^z$ be a sequence of points such that $w_n \to
+\infty$ (i.e. it escapes every compact; recall that $(M,g)$ is
non-total imprisoning thus the past ray $\sigma^z$ escapes every
compact). There are timelike curves $\eta_n$ connecting $w_n$ to
$z_n$ and this curve has no intersection with $J^{-}(z)$ but for the
starting point $w_n$, for otherwise $w_n \in I^{-}(z)$ which is
impossible because $\sigma^z$ is a lightlike ray. Without loss of
generality we can assume that there is $N>0$ such that for $n>N$,
$\eta_n$ intersects the open set
$A_n=B_{n+1}(q)\backslash\{\bar{B}_{n}(q)\cup[ J^{+}(x)\cap
J^{-}(z)]\}$ (just pass to a subsequence and relabel it). Let
$\Omega_n: M \to [1, +\infty)$, be a function  equal to $1$ outside
$A_n$ and sufficiently large on $A_n$ that the Lorentzian length of
$\eta_n$ with respect to $\Omega_n g$ is greater than $n$. Define
$\tilde{g}=\Pi_{n}\Omega_n g$, then $d_{\tilde{g}}(w_n,z_n)>n$, and
thus since $w_n \in J^{+}(x)$, $d_{\tilde{g}}(x,z_n)>n$ which
implies that there is an infinity discontinuity on the Lorentzian
distance $d_{\tilde{g}}$ as $(x,z_n) \to (x,z)$. The contradiction
proves that the spacetime is globally hyperbolic. \hfill$\square$

\end{pf}

\section{Conclusions}
It has been proved that causal simplicity and causal continuity can
be characterized through the continuity properties of the Lorentzian
distance at those pairs of events where it vanishes. The non-total
imprisoning spacetime is causally simple if and only if for every
metric choice in the conformal class the Lorentzian distance is
continuous wherever it vanishes. Similarly, a strongly causal
spacetime is causally continuous if and only if there is at least
one choice of metric in the conformal class such that the Lorentzian
distance is continuous wherever it vanishes. Using some preliminary
lemmas it has also been shown that a non-total imprisoning spacetime
is globally hyperbolic if and only if the Lorentzian distance is
continuous for every choice of metric in the conformal class.

Other results obtained in this work are the characterization of the
distinction condition given by theorem \ref{njs} and that of global
hyperbolicity given by corollary \ref{bzx}.

\section*{Acknowledgments}
This work has been partially supported by GNFM of INDAM.


\end{document}